\documentclass[final, 1p, times]{elsarticle}

\usepackage{graphicx}
\usepackage{amssymb}
\usepackage{amsmath}
\usepackage{amsfonts}
\usepackage{bbm}
\usepackage{braket}
\usepackage{hyperref}

\usepackage{lineno}
\usepackage{xcolor}

\journal{Annals of Physics Special Issue: Localisation 2020}

\begin{document}

\begin{frontmatter}


\title{Localization properties in Lieb lattices and their extensions}



\author[1]{Jie Liu}
\ead{liujie@smail.xtu.edu.cn}
\author[1]{Xiaoyu Mao}
\ead{Maoxiaoyu@smail.xtu.edu.cn}
\author[1]{Jianxin Zhong}
\ead{jxzhong@xtu.edu.cn}
\author[1,2]{Rudolf A.\ R\"{o}mer}
\ead{r.roemer@warwick.ac.uk}

\address[1]{School of Physics and Optoelectronics, Xiangtan University, Xiangtan 411105, China}
\address[2]{Department of Physics, University of Warwick, Coventry, CV4 7AL,
	United Kingdom}

\begin{abstract}
We study the localization properties of generalized, two- and three-dimensional Lieb lattices, $\mathcal{L}_2(n)$ and $\mathcal{L}_3(n)$, $n= 1, 2, 3$ and $4$, at energies corresponding to flat and dispersive bands using the transfer matrix method (TMM) and finite size scaling (FSS). 
We find that the scaling properties of the flat bands are different from scaling in dispersive bands for all $\mathcal{L}_d(n)$. 
For the $d=3$ dimensional case, states are extended for disorders $W$ down to $W=0.01 t$ at the flat bands, indicating that the disorder can lift the degeneracy of the flat bands quickly. 
The phase diagram with periodic boundary condition for $\mathcal{L}_3(1)$ looks similar to the one for hard boundaries \cite{Liu2020LocalizationLattices}. 
We present the critical disorder $W_c$ at energy $E=0$ and find a decreasing $W_c$ for increasing $n$ for $\mathcal{L}_3(n)$, up to $n=3$. 
Last, we show a table of FSS parameters including so-called irrelevant variables; but the results indicate that the accuracy is too low to determine these reliably.
\end{abstract}


\end{frontmatter}


\section{Introduction}

Flat band systems, in which the absence of a dispersion in the  band structure leads to a highly dramatic macroscopic degeneracy, have at the flat band energy an effectively reduced kinetic energy. Hence other terms in the Hamiltonian can become prominent, such as many-body interactions. This mechanism leads to a convenient construction of various platform for studying many-body physics, such as the fractional quantum Hall effect \cite{Tang2011,Neupert2011,Sun2011NearlyTopology}, spin liquids \cite{Savary2017,Balents2010}, ferromagnetism \cite{Mielke1993FerromagnetismModel,Tasaki1998FromModel}, and superconductivity \cite{Miyahara2007BCSLattice,Julku2016GeometricBand,Kopnin2011}. In recent years, artificial lattices \cite{Leykam2018}, for instance, in photonic  \cite{Mukherjee2015a,Vicencio2015a,Guzman-Silva2014,Diebel2016,Taie2015,Nixon2013} and cold atom systems \cite{Shen2010SingleLattices,Goldman2011,Apaja2010}, allow to realize experimentally also the probing of the \emph{novel} many-body problems.

As is well known, states in a flat band are localized \cite{Vicencio2015a} because of the high degeneracy. Hence disorder, which one should expect to destroy this degeneracy, might also, at least initially, destroy the localization. So what will happen after disorder is being included in the Hamiltonians describing these localized flat bands is an interesting question.

\begin{figure}[]
    \centering
    (a)\includegraphics[width=0.20\columnwidth]{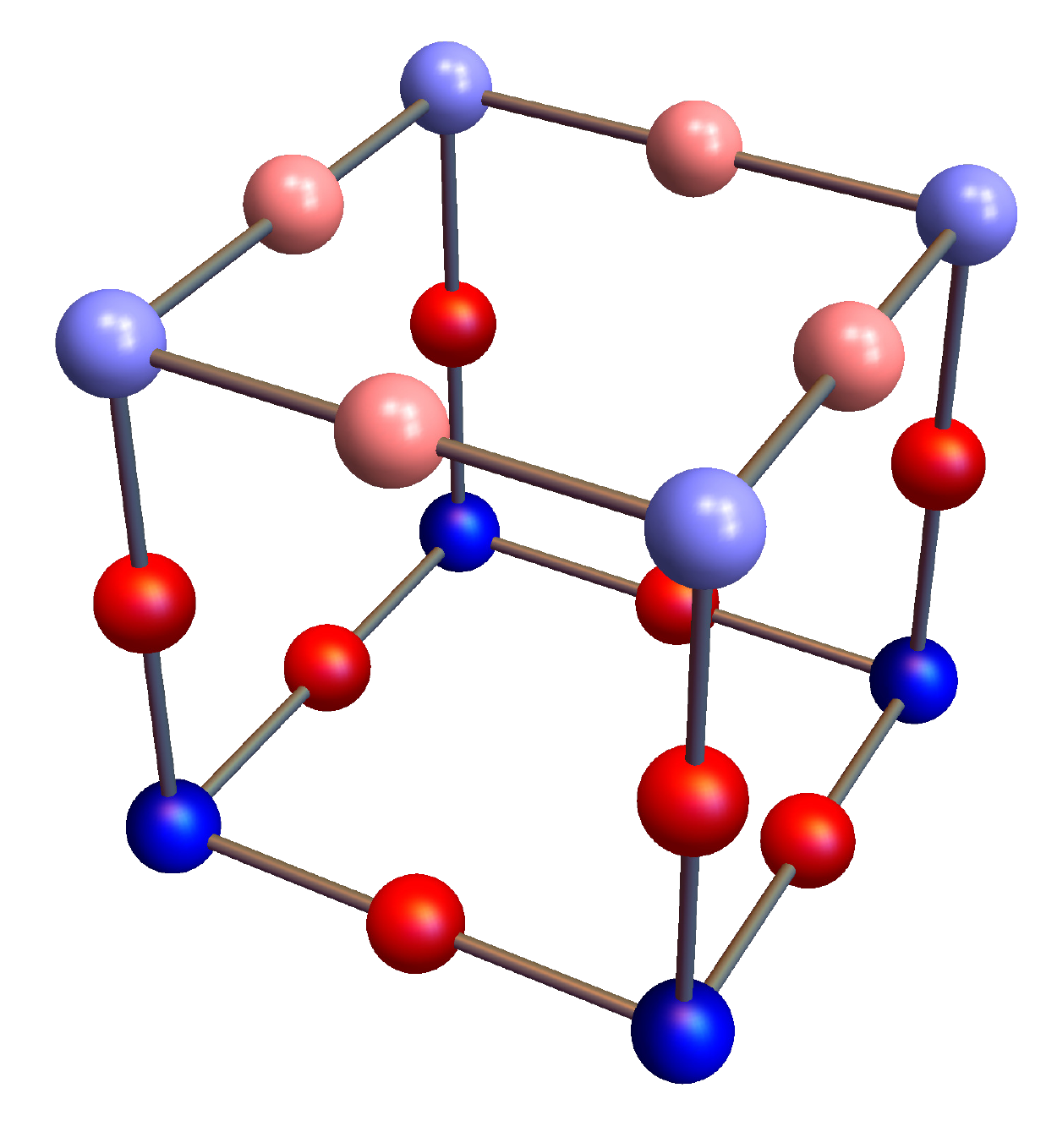}
    (b)\includegraphics[width=0.20\columnwidth]{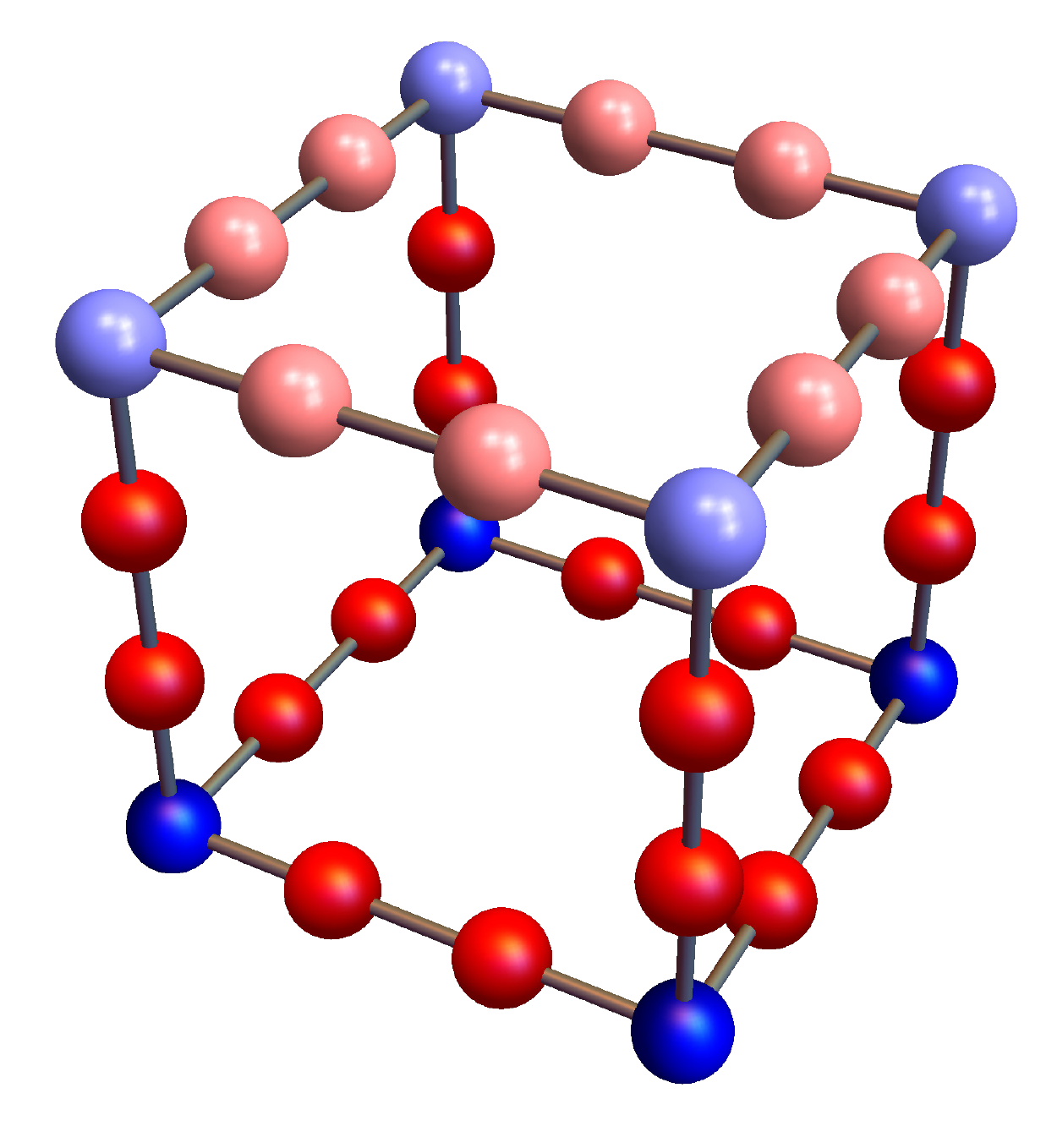}
    (c)\includegraphics[width=0.20\columnwidth]{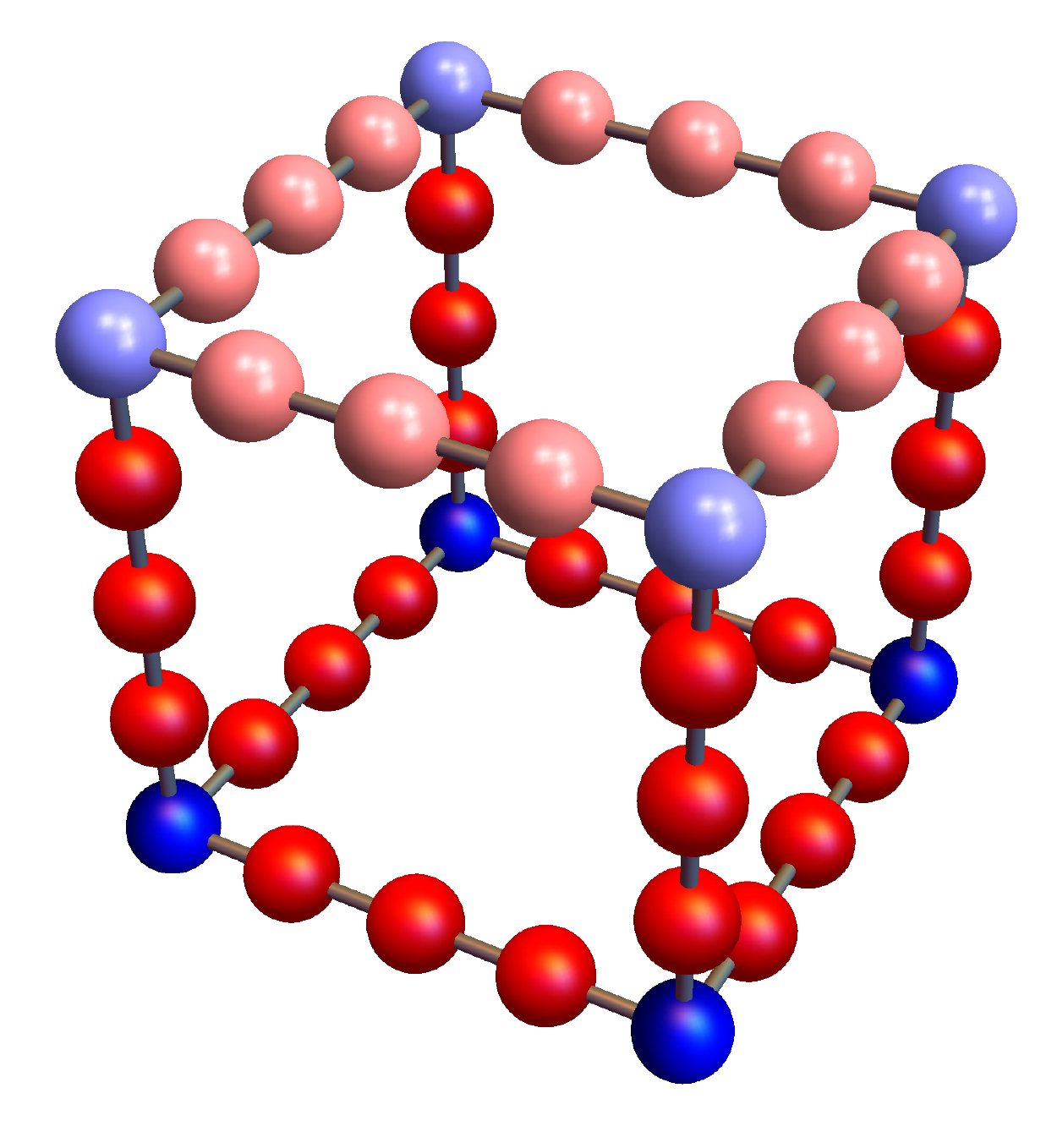}
    (d)\includegraphics[width=0.20\columnwidth]{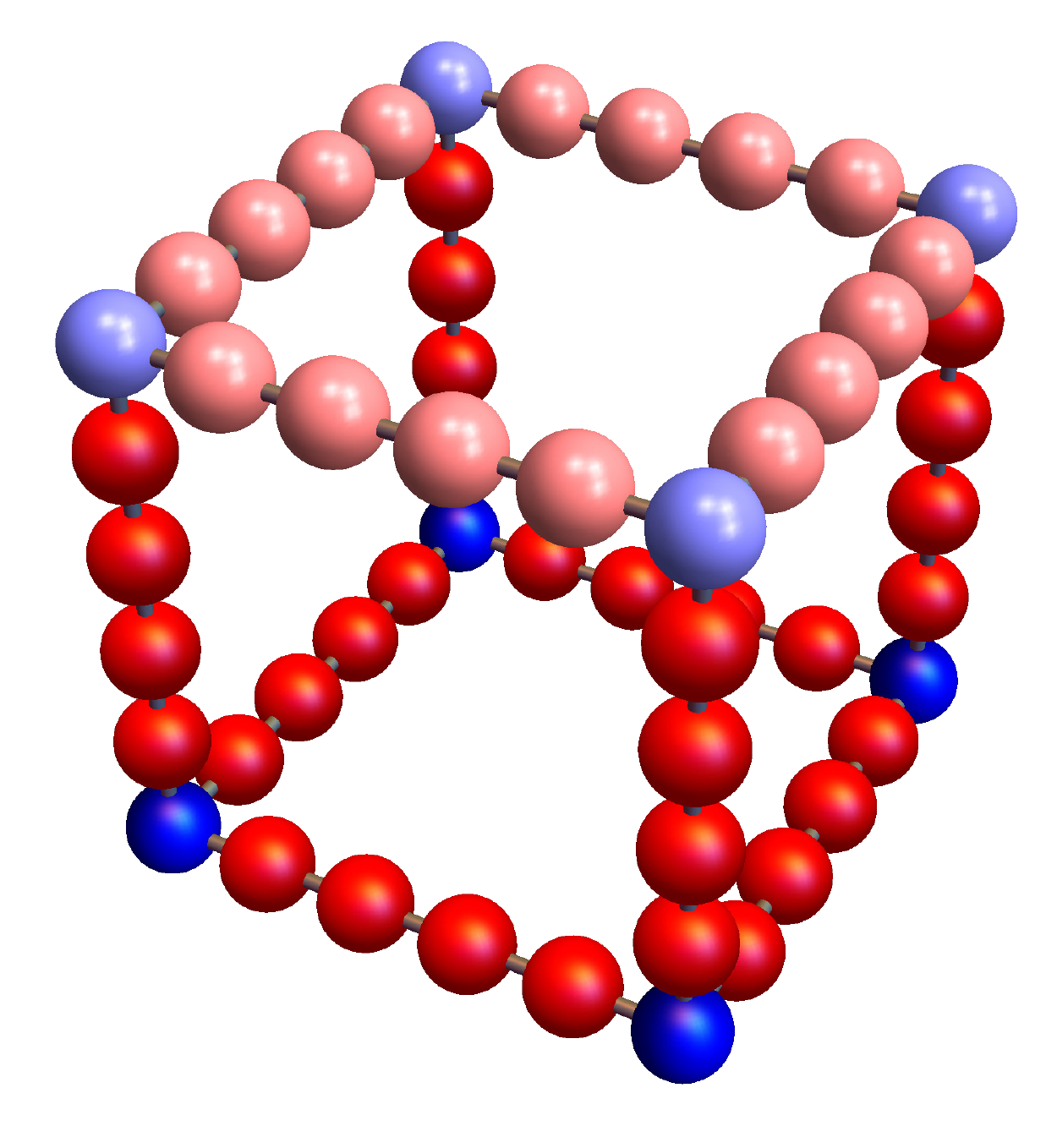}
    \caption{Schematic representation of Lieb and extended Lieb lattices $\mathcal{L}_d(n)$. The lightly coloured spheres highlight the situation in 2D while together with the fully coloured spheres they represent the 3D lattices. Blue spheres denote undecimated lattices site while the red sphere will be decimated in the TMM formulation. The dark lines are guides to the eye only. The labels (a), (b), (c) and (d) denote $\mathcal{L}_2(1)$/$\mathcal{L}_3(1)$, $\mathcal{L}_2(2)$/$\mathcal{L}_3(2)$, $\mathcal{L}_2(3)$/$\mathcal{L}_3(3)$ and $\mathcal{L}_2(4)$/$\mathcal{L}_3(4)$, respectively.}
    \label{fig:Lieb_schematic}
\end{figure}

Flat band system can be constructed in many models \cite{Tang2011,Neupert2011,Sun2011NearlyTopology,Tasaki1998FromModel,Weeks2010TopologicalLattices}. The Lieb lattice \cite{Julku2016GeometricBand,Lieb1989TwoModel,Qiu2016DesigningSurface} is one of the simplest and most famous two-dimensional flat band system. Actually, the $CuO_2$ plane of cuprate superconductors is also a Lieb lattice, namely $\mathcal{L}_2(1)$ in our notation. It contains three atoms per unit cell as shown in Fig.\ \ref{fig:Lieb_schematic}(a). In the figure, we also introduce its extensions $\mathcal{L}_2(n)$, $n=2, 3$ and $4$, shown in panels (b)-(d), respectively. The three-dimensional Lieb lattice and its extensions are also shown in Fig.\ \ref{fig:Lieb_schematic}(a)-(d). The number of flat bands is related to the number of central red atoms between two nearest blue atoms. In short, for $\mathcal{L}_d(n)$, the number of flat band is $n$ with $d-1$ degeneracy. The central energy at $E=0$ is part of a flat band for $\mathcal{L}_2(1)$, $\mathcal{L}_2(3)$, $\mathcal{L}_3(1)$ and $\mathcal{L}_3(3)$ while it remains in a dispersive band for $\mathcal{L}_2(2)$, $\mathcal{L}_2(4)$, $\mathcal{L}_3(2)$ and $\mathcal{L}_3(4)$\cite{Mao2020b,Liu2020LocalizationLattices}.

This paper is organized as follows. In section \ref{sec-2d} and section \ref{sec-3d} we will discuss the two-dimensional and three-dimensional Lieb lattices and their extensions, respectively. The conclusion is given in section \ref{sec-conclusion}. 

\section{The two-dimensional Lieb lattice and its extensions}
\label{sec-2d}

\subsection{Previous results for disordered $\mathcal{L}_2(n)$ lattices}


In our previous paper \cite{Mao2020b}, we study the localization properties of $\mathcal{L}_2(n)$ in detail. 
Using direct diagonalization for small system sizes, we compute the density of states (DOS). We can see that in the presence of disorder, the interplay between flat bands and dispersive bands is prominent for all $\mathcal{L}_2(n)$, $n=1,2,3$ and $4$. The disorder can quickly lift the degeneracy of flat bands and make the states merge with the neighboring dispersive bands. When $W\gtrsim 2$, the flat band DOS loses its peaks distinguishing and becomes part of a large bulk DOS.

Next, we use a renormalized transfer-matrix method (TMM) and compute the reduced localization lengths $\Lambda_M(E,W)=\lambda(E,W)/M$, where $M$ corresponds to the width of the quasi-1D transfer-matrix strip.
For all $\mathcal{L}_d(n)$, we find that for $W\gtrsim t$ all states are localized. The localization lengths for states at flat band energies are about one order of magnitude smaller than for states from dispersive bands. We employ traditional one-parameter finite-size scaling methods to estimate a scaling parameter $\xi(E,W)$ \cite{MacKinnon1981One-ParameterSystems}. The $\Lambda(E,W)/M$ can be described after scaling by a single scaling branch, corresponding to a fully localized behaviour. After fitting the $\xi$ with disorder $W$, we use three fitting forms, power low form as $\xi(W) \propto W^{-2}$\cite{MacKinnon1983a}, a non-universal form $\xi(W)=a W^{-\alpha}\exp(\beta W^{-\gamma})$ and a constraint form $\xi(W)=a W^{-2}\exp(\beta W^{-1})$. For $\mathcal{L}_2(1)$ and $\mathcal{L}_2(3)$ at the flat band energy $E=0$, we find that the usual power low form $\xi(W) \propto W^{-2}$ for 1D localization fits well for disorder around $1<W<2$. However, for $\mathcal{L}_2(2)$ and $\mathcal{L}_2(4)$ at $E=0$, which is an energy in a dispersive band for these lattices, none of the fits gives a convincing result.

\subsection{Scaling function $\Lambda_M$ vs reduced correlation length $\xi_M$ for $\mathcal{L}_2(n)$}

\begin{figure}[tb]
    \centering
    (a)\includegraphics[width=0.9\columnwidth]{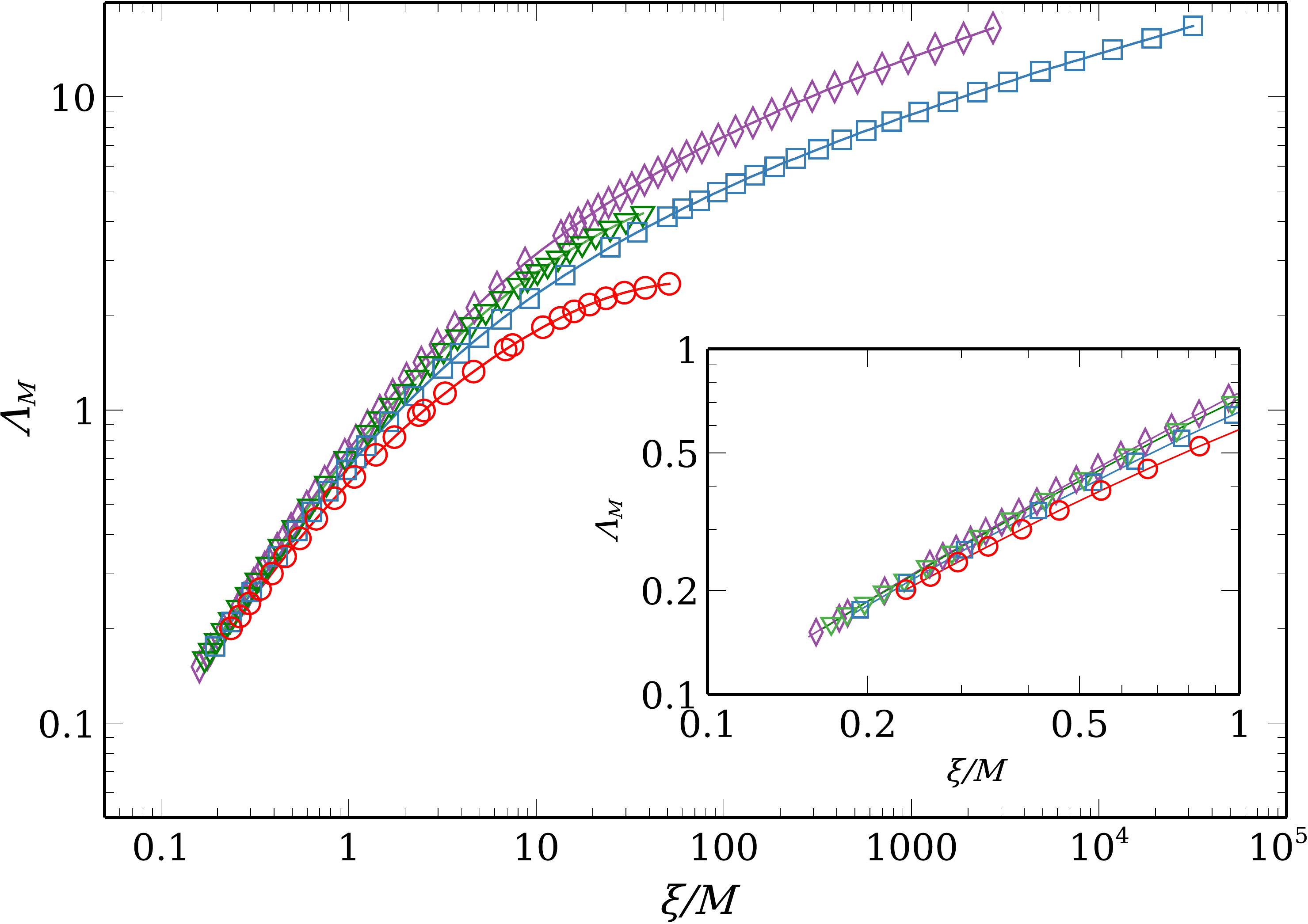}
    (b)\includegraphics[width=0.9\columnwidth]{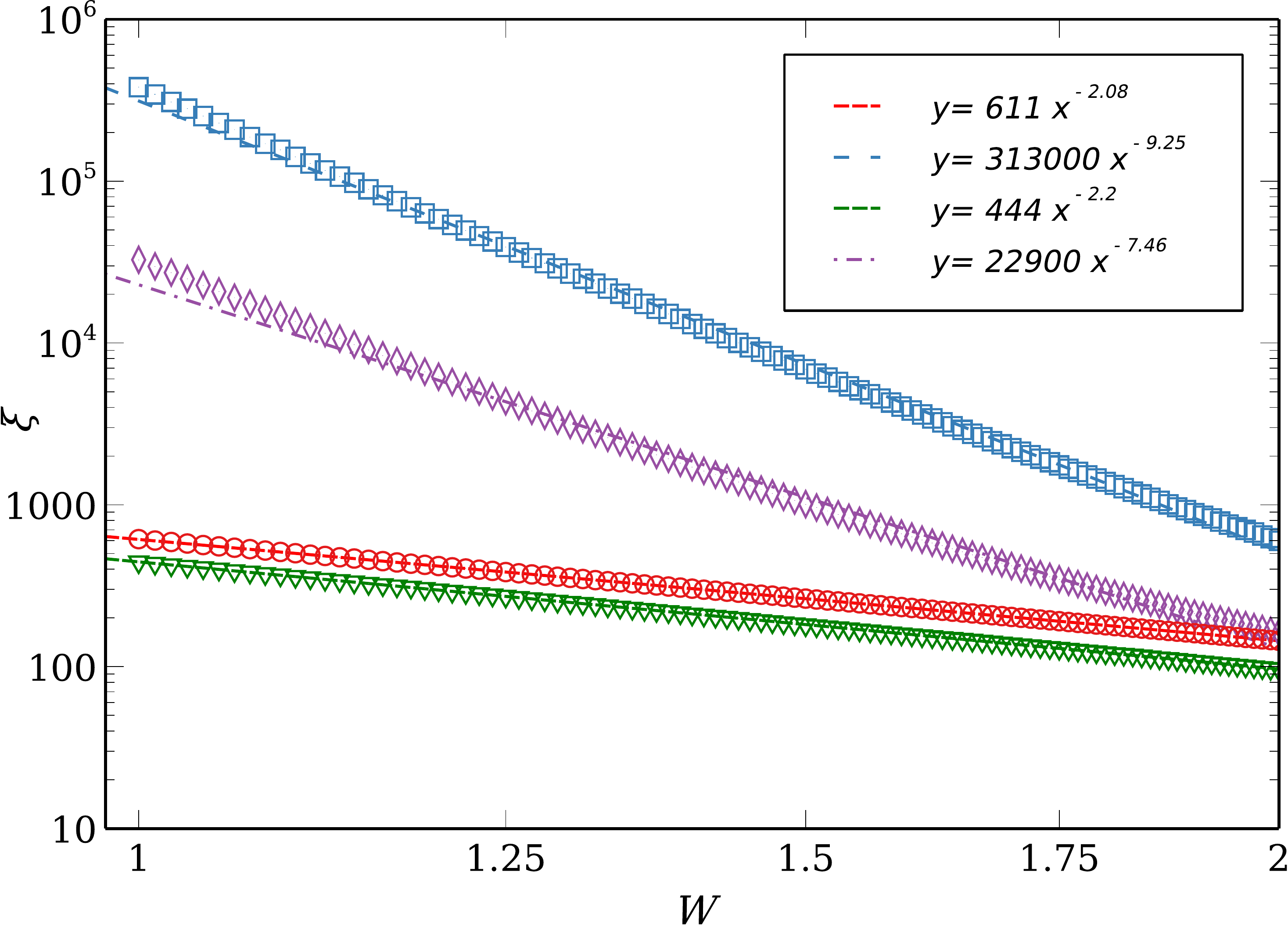}
    \caption{(a) Scaled localization length $\Lambda_M(0,W)$ versus $\xi/M$ at energy $E=0$ for $\mathcal{L}_2(1)$ (red $\bigcirc$), $\mathcal{L}_2(2)$ (blue $\square$), $\mathcal{L}_2(3)$ (green $\triangledown$) and $\mathcal{L}_2(4)$ (purple $\diamondsuit$). For clarity, lines show all data points while symbols denote only about $15\%$ of all data.
    Inset: the detail of small reduced correlation length. 
    %
    (b) Scaling parameters $\xi(0,W)$ for the same Lieb lattices as in (a). The dashed lines represent the power law fit functions $a x^b$.
    Error bars are within symbol size in both panels.}
    \label{fig:Scaling_Func_L2x}
\end{figure}

The scaled localization lengths $\Lambda_M(0,W)$ as a function of scaled correlation length $\xi/M$ for $\mathcal{L}_2(n)$, $n=1, 2, 3$ and $4$ are shown in Fig.\ \ref{fig:Scaling_Func_L2x}(a) at energy $E=0$. This corresponds to flat bands for $\mathcal{L}_2(1)$ and $\mathcal{L}_2(3)$ and dispersive band for $\mathcal{L}_2(2)$ and $\mathcal{L}_2(4)$. The $\Lambda_M(0,W)$ data all show the behaviour for localized states, scaling as $\Lambda_M(0,W)\propto \xi(0,W)/M$ for large system sizes and large disorders. In this regime, the behaviour of states in the flat bands and in the dispersive bands is similar as shown also in the inset graph of said figure. 

\subsection{Scaling parameter $\xi$ vs disorder $W$ for $\mathcal{L}_2(n)$}


The disorder dependence of the scaling parameter $\xi$ for small disorders $t \leq W \leq 2 t$ is shown in Fig.\ \ref{fig:Scaling_Func_L2x}(b), computed from the $\Lambda_M(E=0,W)$ data of Fig.\ \ref{fig:Scaling_Func_L2x}(a).
We see that the behaviour of $\xi$ for $\mathcal{L}_2(1)$ is comparable to  $\mathcal{L}_2(3)$; both are well-described by a power law $a x^b$ with exponent approximately $b \sim -2$, similar to localization properties of a standard 1D Anderson model \cite{MacKinnon1983a}. This might suggest that the localization behaviour of these flat band states at least for weak disorder is similar to the 1D behaviour.
On the other hand, for the dispersive states of $\mathcal{L}_2(2)$ and $\mathcal{L}_2(4)$ at $E=0$, we find that the values of $\xi$ are orders of magnitude larger than for $\mathcal{L}_2(1)$ and $\mathcal{L}_2(3)$. The simple power-law also does not fit anymore and we rather see the more standard behaviour of a 2D Anderson model \cite{MacKinnon1983a} with a quickly diverging $\xi$ when $W\rightarrow 0$. 
Nevertheless, for both flat band and dispersive band energies, the fits are not very robust and have rather small $p$ values of $< 10^{-10}$. This shows that the true form of the behaviour of $\xi (W \rightarrow 0)$ is yet to be determined.

\subsection{Density of states without Gaussian broadening for $\mathcal{L}_2(n)$}

\begin{figure}[tb]
    \centering
     \quad $\mathcal{L}_2(n)$ \quad\quad \hfil \quad $\mathcal{L}_3(n)$ \quad \\[1ex]
    (a)\includegraphics[width=0.45\columnwidth]{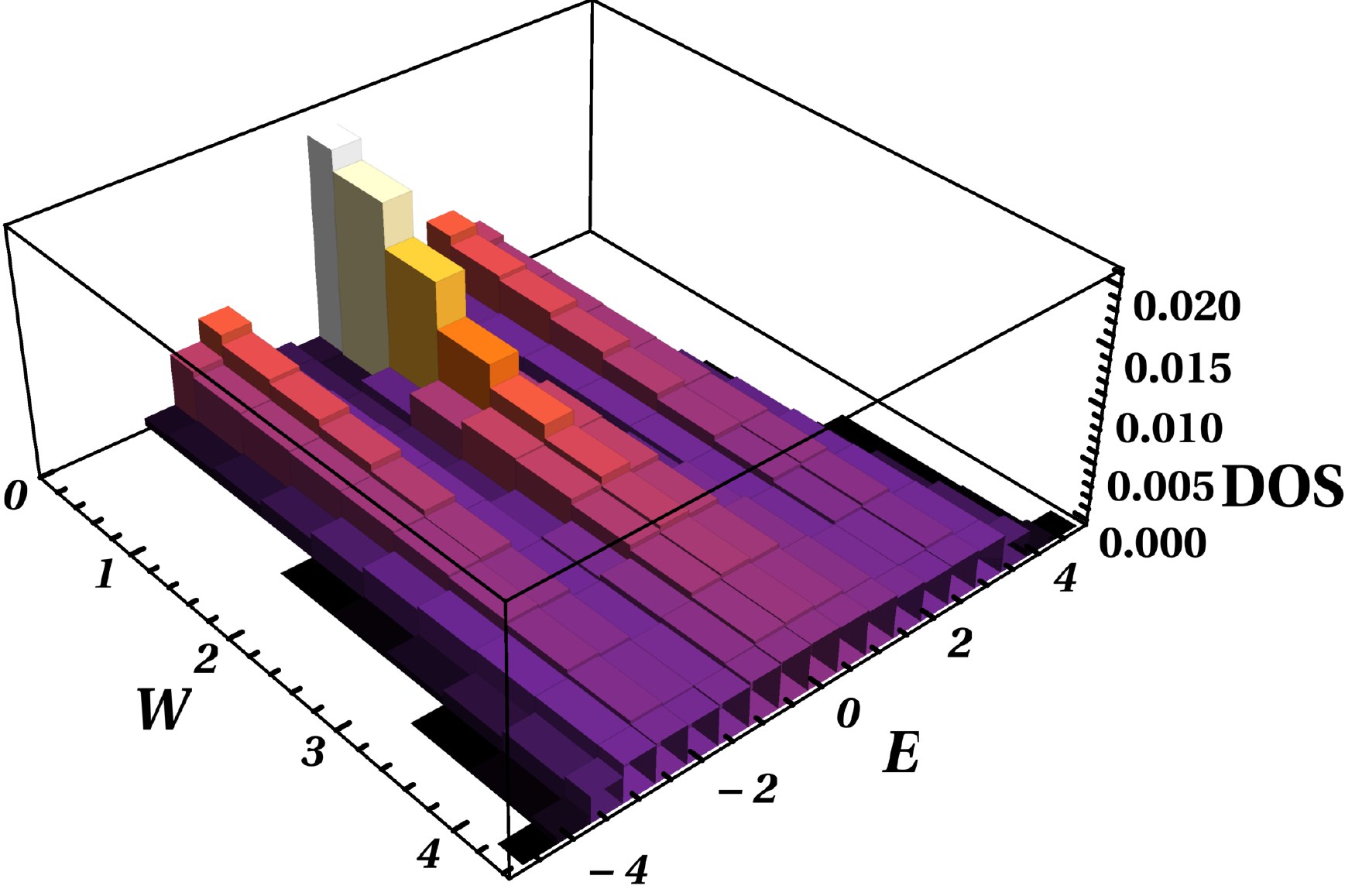}
    \includegraphics[width=0.45\columnwidth]{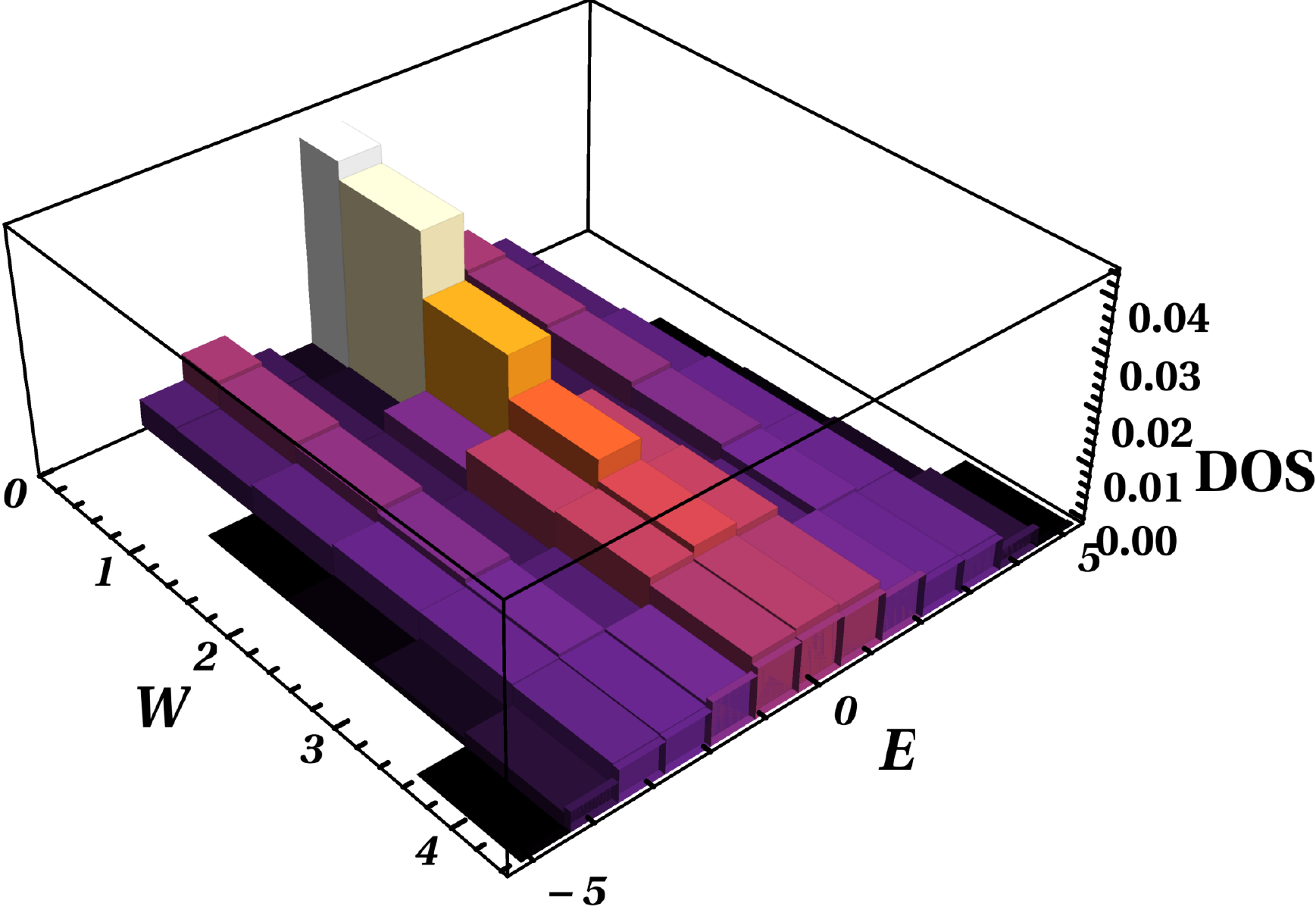}\\
    (b)\includegraphics[width=0.45\columnwidth]{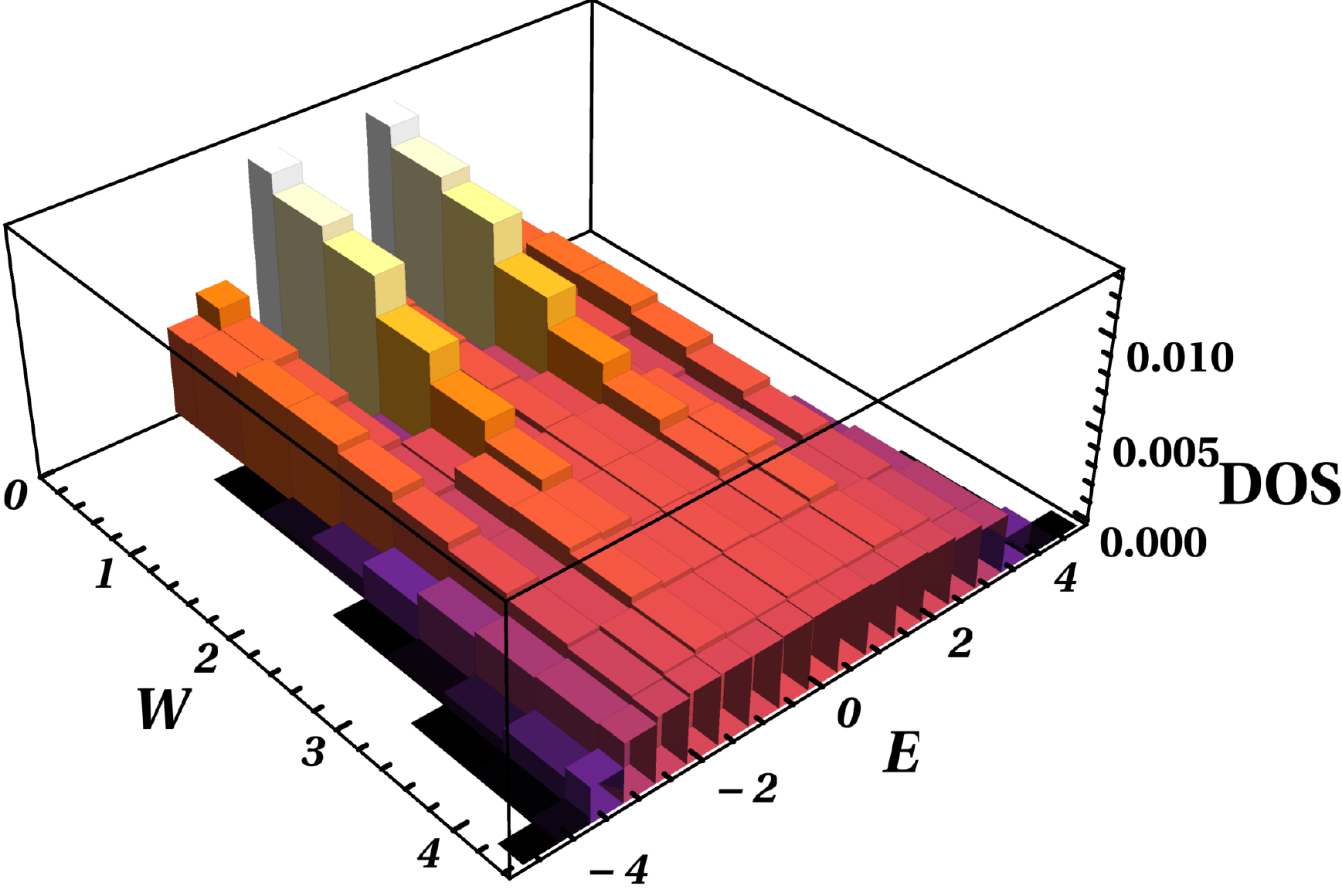}
    \includegraphics[width=0.45\columnwidth]{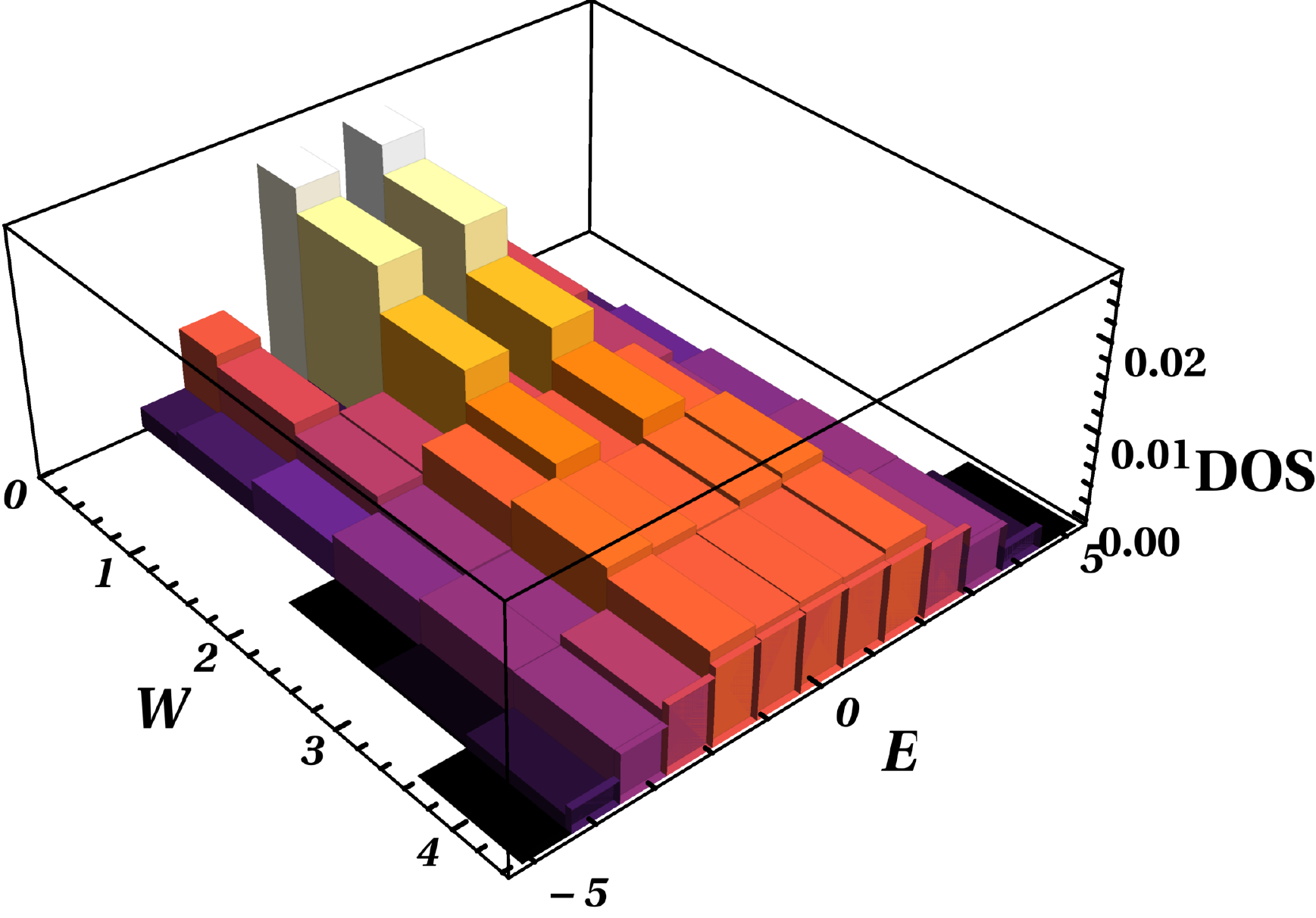}\\
    (c)\includegraphics[width=0.45\columnwidth]{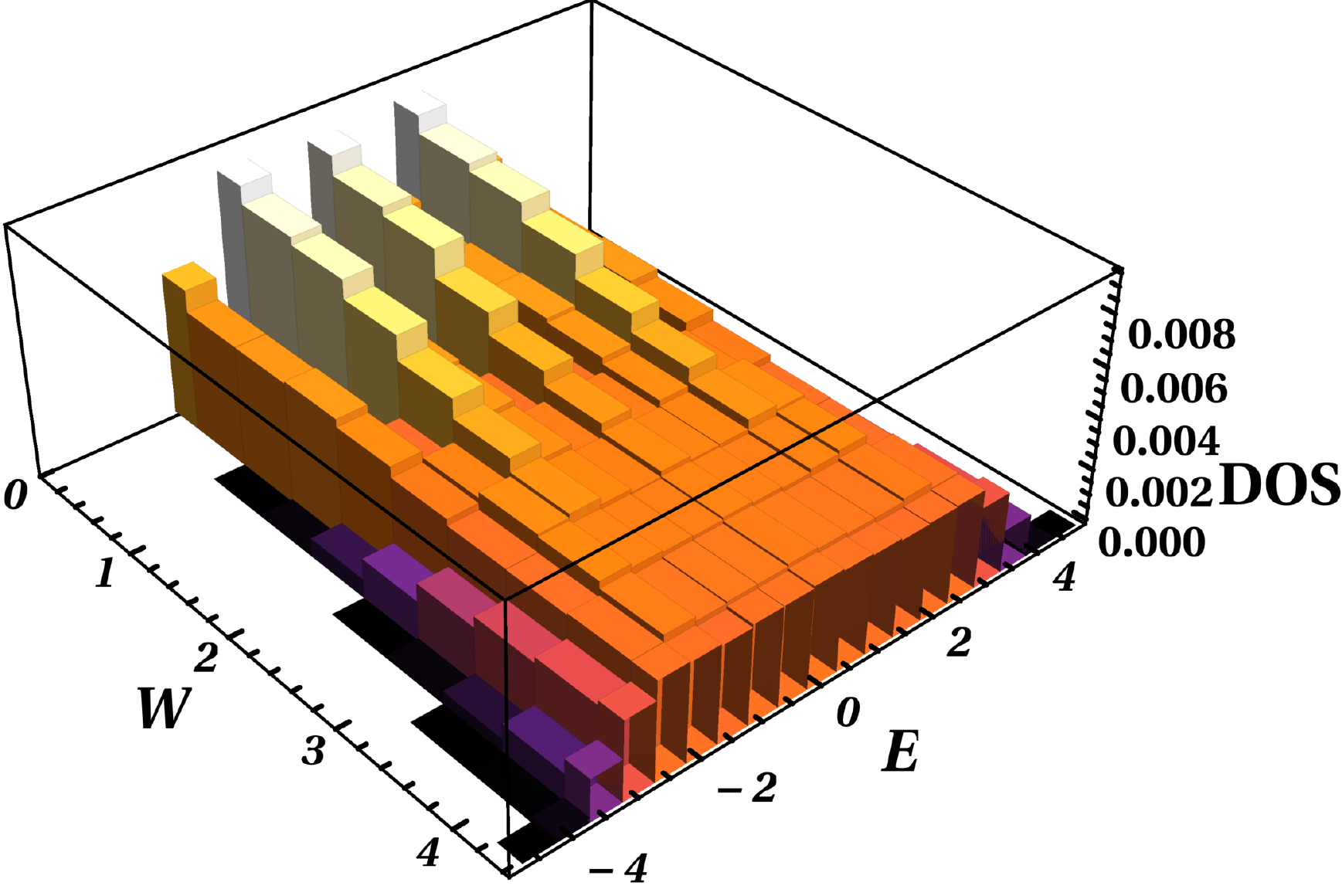}
    \includegraphics[width=0.45\columnwidth]{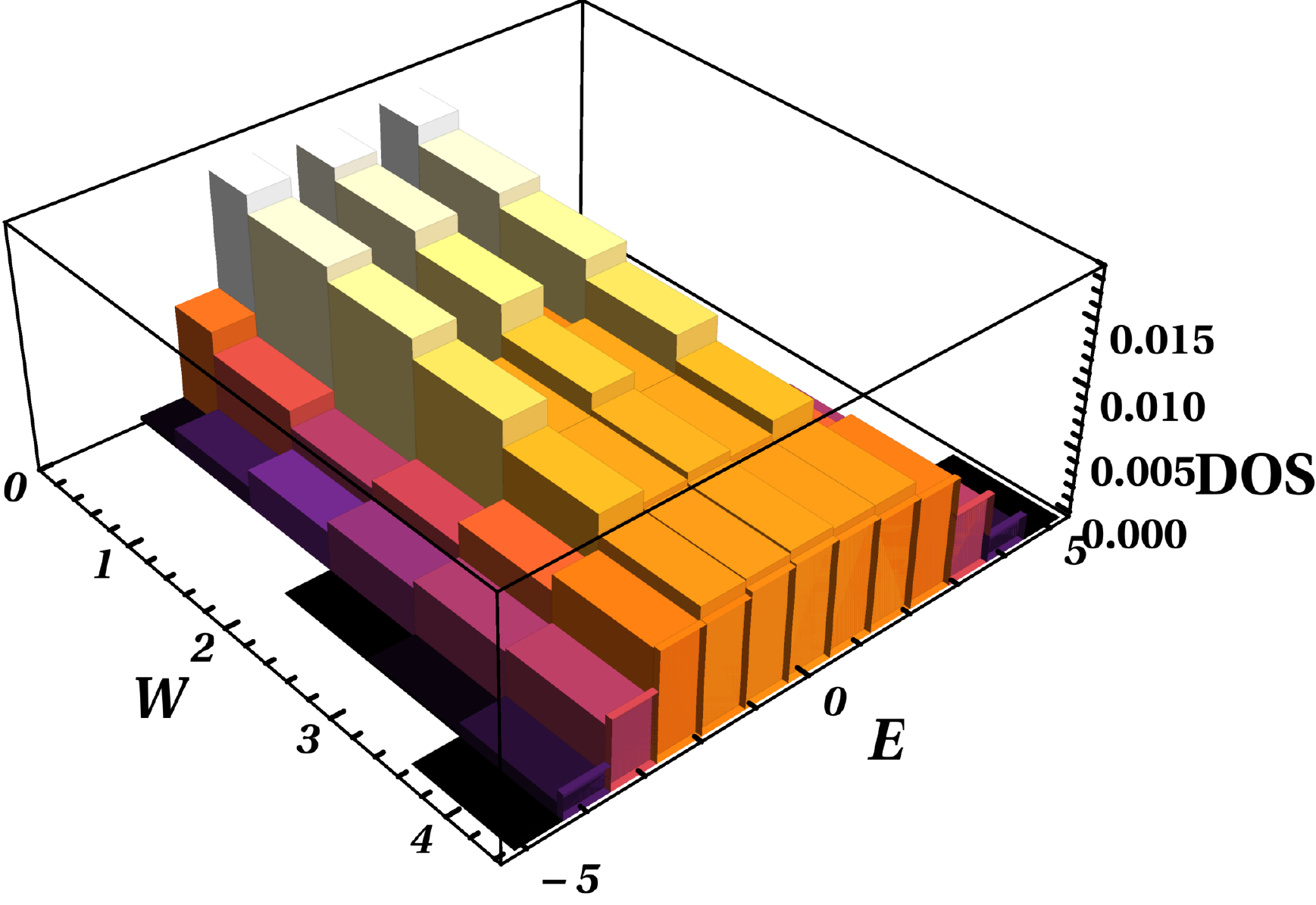}\\
    (d)\includegraphics[width=0.45\columnwidth]{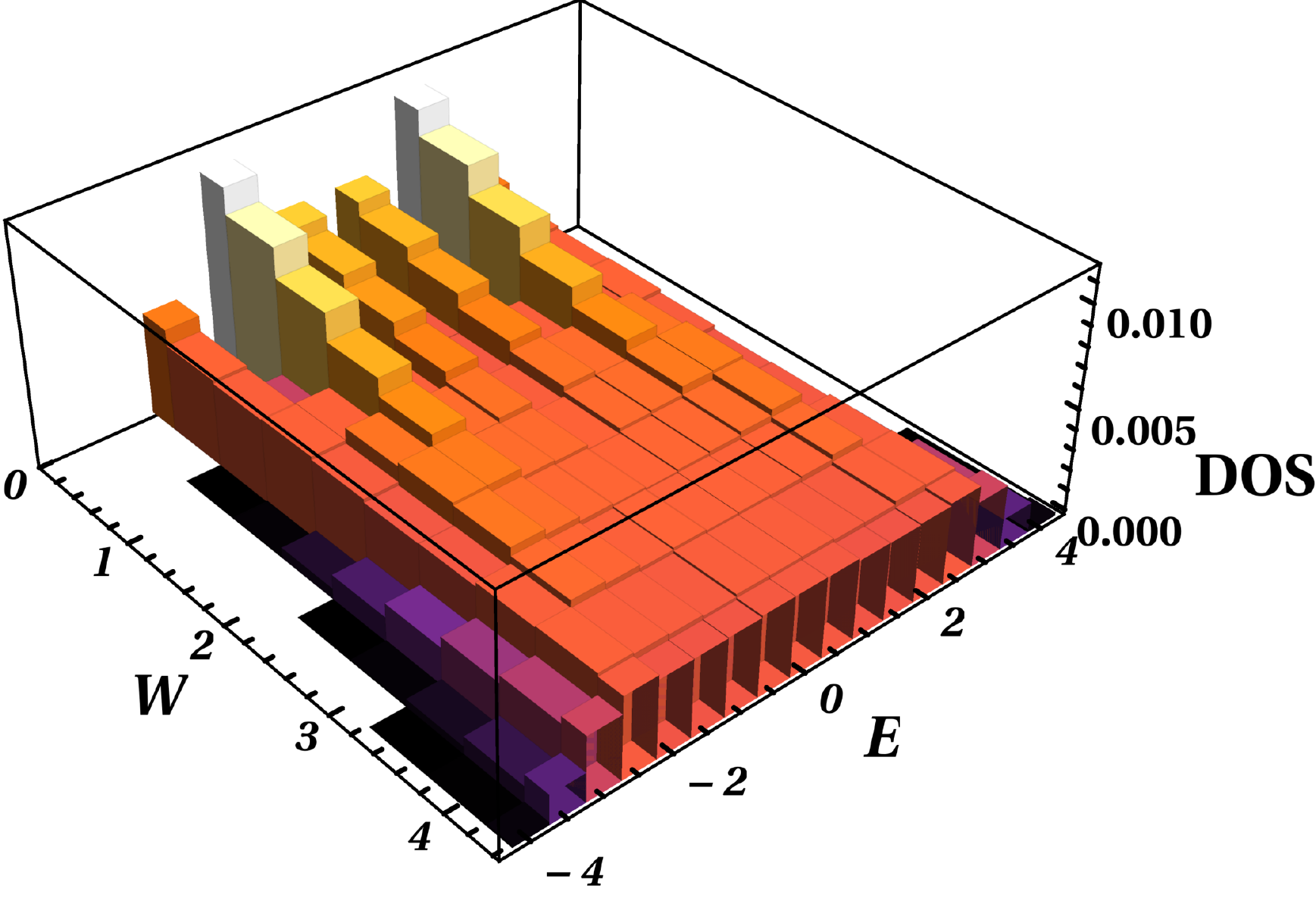}
    \includegraphics[width=0.45\columnwidth]{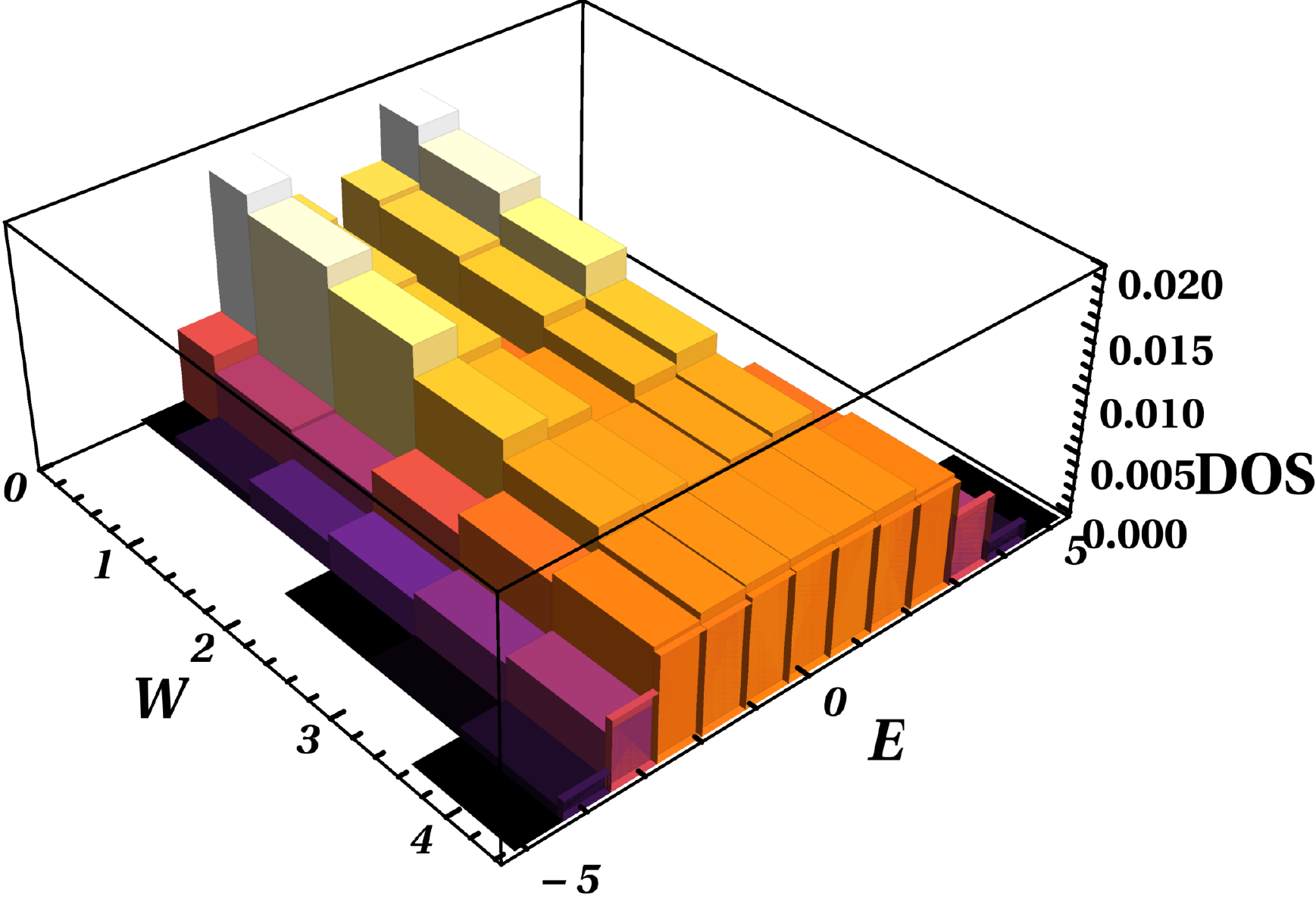}
%
    %
    \caption{
    Normalized bar chart histograms of the $(E,W)$-dependence of the density of states (DOS) for (a) $\mathcal{L}_2(1)$ and $\mathcal{L}_3(1)$, 
    (b) $\mathcal{L}_2(2)$ and $\mathcal{L}_3(2)$, 
    (c) $\mathcal{L}_2(3)$ and $\mathcal{L}_3(3)$, and
    (d) $\mathcal{L}_2(4)$ and $\mathcal{L}_3(4)$. The colors denote different DOS values ranging from zero (deep purple) to maximal (white). Bin widths in $(E,W)$ directions have been chosen for representational clarity.}
    \label{fig:L2x DOS without Gaussian smoothing}
    \label{fig:DOS without Gaussian smoothing}
\end{figure}
The results of the disorder-averaged density of states (DOS), calculated with direct diagonalization, are shown in Fig.\ \ref{fig:L2x DOS without Gaussian smoothing}. The system sizes are $M^2=13^2$, $10^2$, $9^2$, $8^2$ for $\mathcal{L}_2(n)$, $n=1, 2, 3$ and $4$, respectively. The disorder ranges from $W=0$ to $W=5.2$ in step of $0.05$ using $300$ independent random samples. We can see the prominent peaks of flat bands close to $W=0$ have been largely vanished when the disorder reaches up to $W=2$ for all $\mathcal{L}_2(n)$, $n= 1, 2, 3$ and $4$. 

\section{Three-dimensional Lieb lattice and its extensions}
\label{sec-3d}

\subsection{Previous results for disordered $\mathcal{L}_3(n)$ lattices}


In Ref.\ \cite{Liu2020LocalizationLattices}, we investigated the DOS, the localization properties and the phase diagrams for the 3D Lieb lattices $\mathcal{L}_3(n)$, $n=1, 2, 3$ and $4$ as shown in Fig.\ \ref{fig:Lieb_schematic}. Obviously, the main difference to the 2D case is the existence of an Anderson metal insulator transition (MIT) in 3D Lieb lattices \cite{Liu2020LocalizationLattices}. Details of the TMM construction for $\mathcal{L}_3(n)$ can also be found in  Ref.\ \cite{Liu2020LocalizationLattices} along with finite-size scaling results for the universal critical exponent $\nu$ of the localization lengths. We found $\nu$ to be in good agreement with the currently accepted value of $\nu=1.590 (1.579, 1.602)$ \cite{Slevin1999b,Rodriguez2011MultifractalTransition} for the Anderson transition. In the following, we shall elaborate on the stability of the phase diagrams to a change in boundary conditions, highlight the various positions of the transitions in the phase diagrams, comment on the possibility of FSS with irrelevant scaling parameters and also provide the DOS without Gaussian-broadening.

\subsection{Phase diagram with periodic boundary condition for $\mathcal{L}_3(1)$}

\begin{figure}[tb]
    \centering
    (a)\includegraphics[width=0.45\columnwidth]{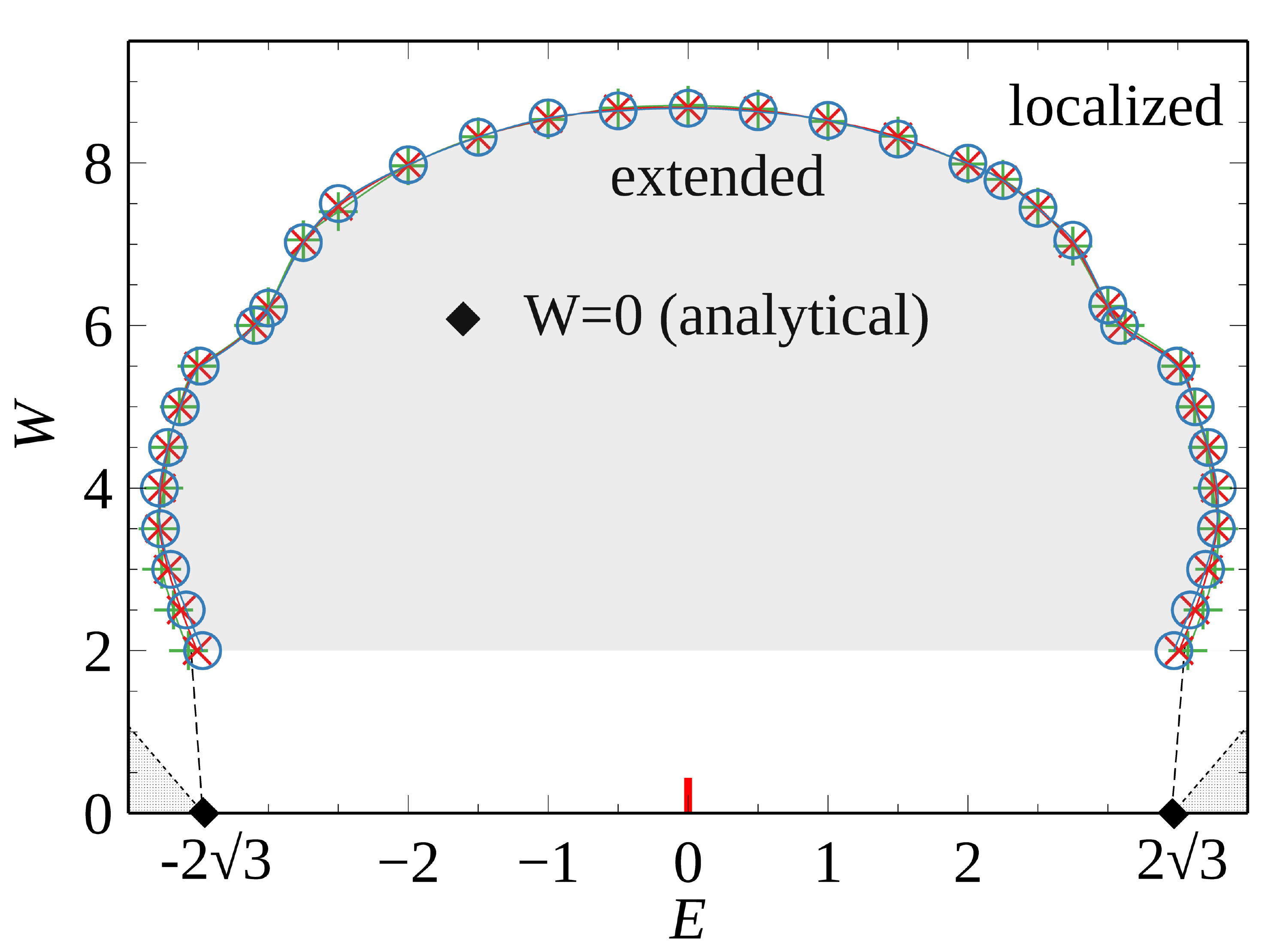}
    (b)\includegraphics[width=0.45\columnwidth]{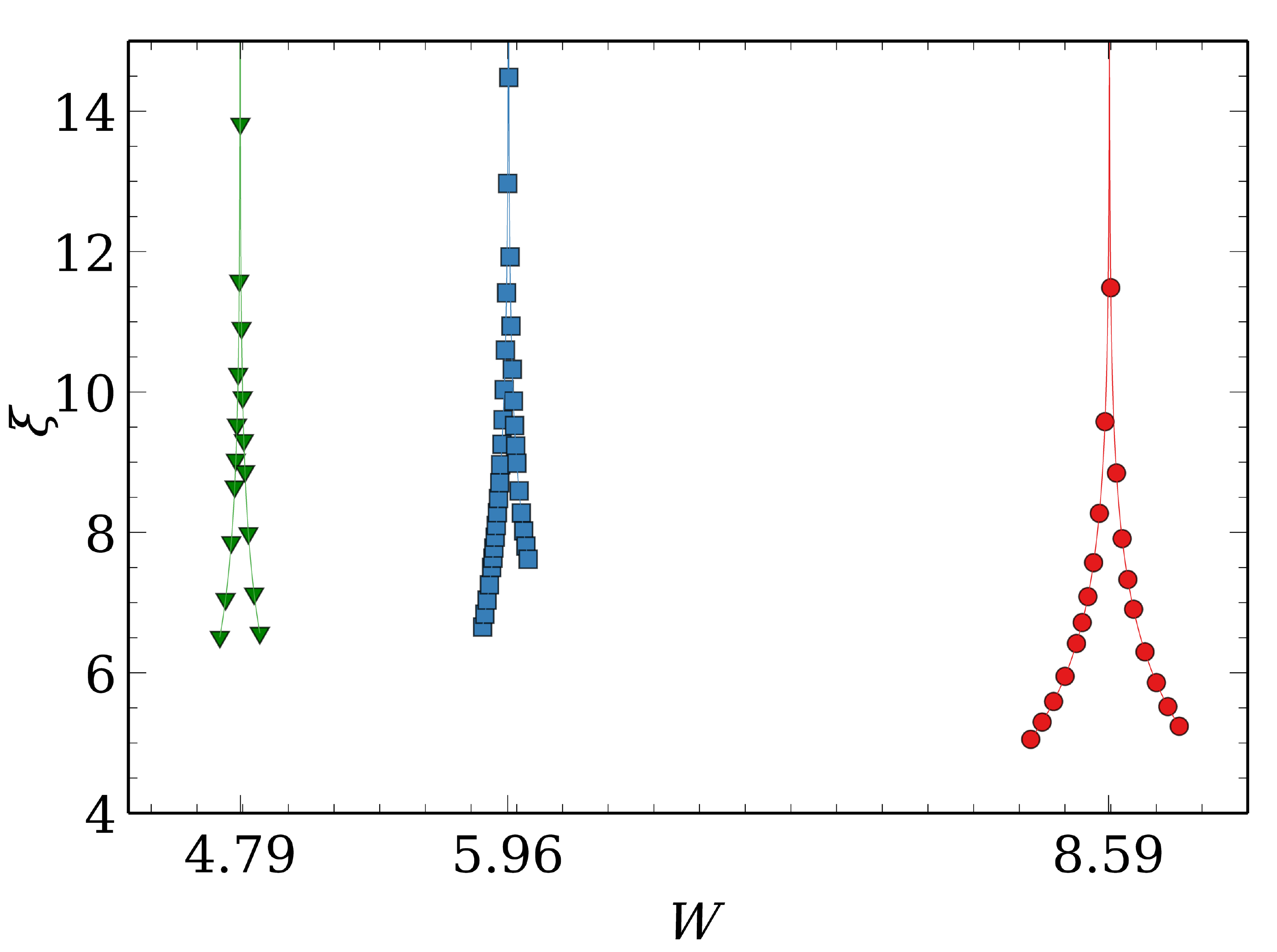}
    \caption{
    (a) Phase diagram for $\mathcal{L}_3(1)$ in case of periodic boundaries. The three solid and colored lines represent the approximate location of the phase boundary estimated from small $M$, i.e.\ the blue line/$\bigcirc$ comes from widths $M=4$ and $M=6$, the red line/$\times$ from $M=4$ and $M=8$, and the green line/$+$ from $M=6$ and $M=8$. 
    The shaded area in the center contains extended states while states outside the phase boundary are localized. The dashed lines on both sides are guides-to-the-eye for the expected continuation of the phase boundary for $W<2$.
    The red short vertical line at $E=0$ represents the position of the doubly-degenerate flat band. 
    The diamonds ($\blacklozenge$) denote the band edges for $W=0$, i.e.\ $E_\mathrm{min}=-2\sqrt{3}$ and $E_\mathrm{max}=2\sqrt{3}$. The dotted lines are the theoretical band edges $\pm \left( |E_\mathrm{min}| + W/2 \right)$ and the forbidden areas below those band edges have been shaded.
    (b) Scaling parameters $\xi$ versus disorder $W$ for $\mathcal{L}_3(1) $(red $\bigcirc$), $\mathcal{L}_3(2)$ (blue $\square$) and $\mathcal{L}_3(3)$ (green $\triangledown$) at energy $E=0$. The expansion parameters $n_r$, $n_i$, $m_r$ and $m_i$ are the same as the highlighted line in Table \ref{table:critical parameters for L3x}. 
    }
    \label{fig:Periodic_phase_diagram_L31}
    \label{fig:Scaling_Parameter_L3x}
\end{figure}

The phase diagrams given in Ref.\ \cite{Liu2020LocalizationLattices} have been computed for the $\mathcal{L}_3(n)$ lattices with hard wall boundaries. In Fig.\  \ref{fig:Periodic_phase_diagram_L31}(a) we now show a phase diagram for $\mathcal{L}_3(1)$ with periodic boundary condition. The phases have been determined by the scaling behaviour of the $\Lambda_M(E,W)$ for system sizes $M=4$, $M=6$ and $M=8$ with error $\leq 0.1\%$\cite{EILMES2008}. Comparing this to the results obtained with hard boundary conditions, we find that it looks very similar as expected although the extended region is a little wider in the $E$ axis. As for the hard wall case, we can identify a re-entrant region around disorder $W=4$ and a shoulder at $E\sim 3$ and $W\sim 6$. Hence as expected, the change in boundary conditions does not change the phase diagrams appreciably already with the small system sizes and modest disorder averages as used here and in Ref.\ \cite{Liu2020LocalizationLattices}.  

\subsection{Localization and extended transition with 0.01 $\leq$ $W$ $\leq$ 2.0 }

\begin{figure}[tb]
    \centering
    (a)\includegraphics[width=0.45\columnwidth]{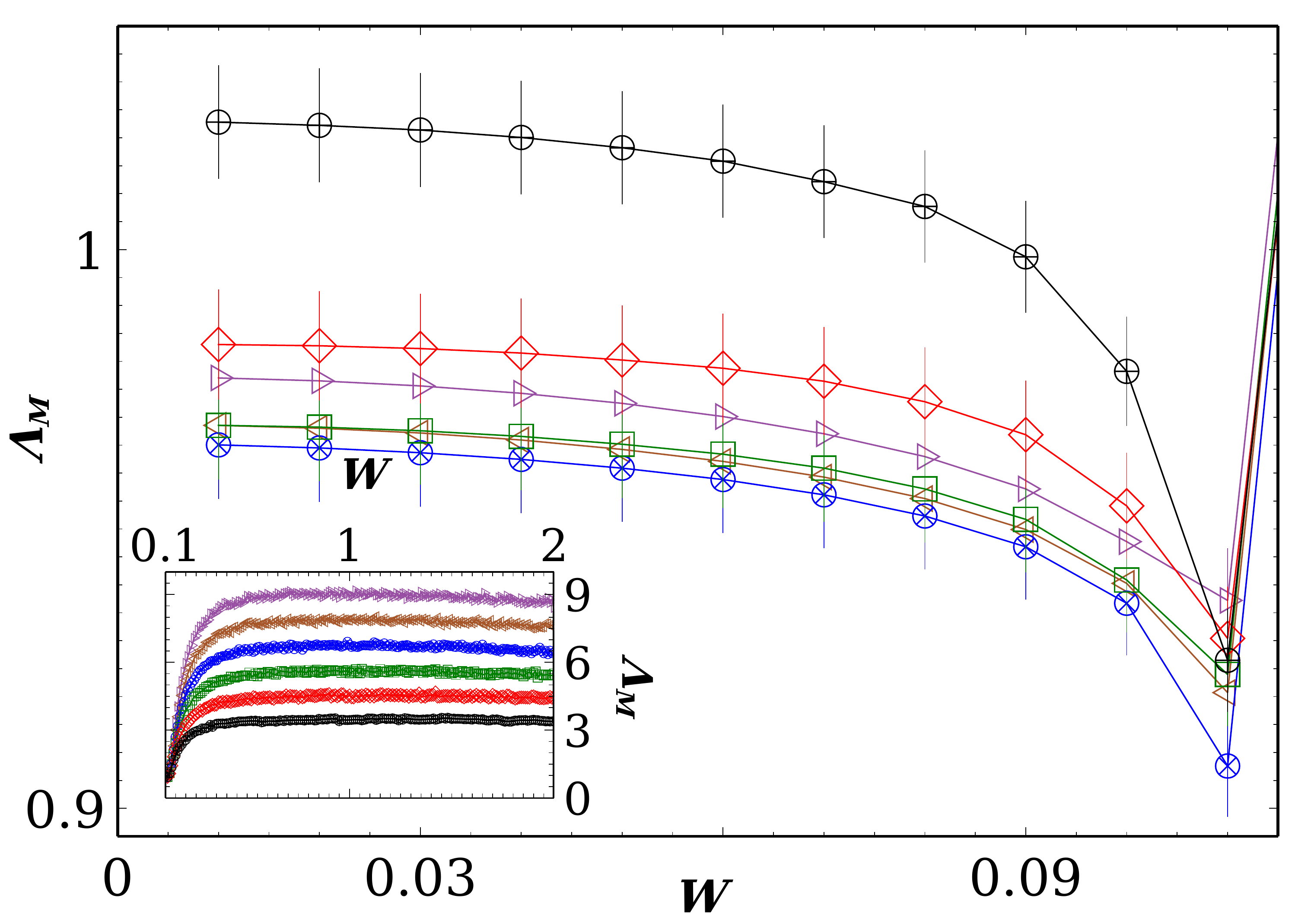}
    (b)\includegraphics[width=0.45\columnwidth]{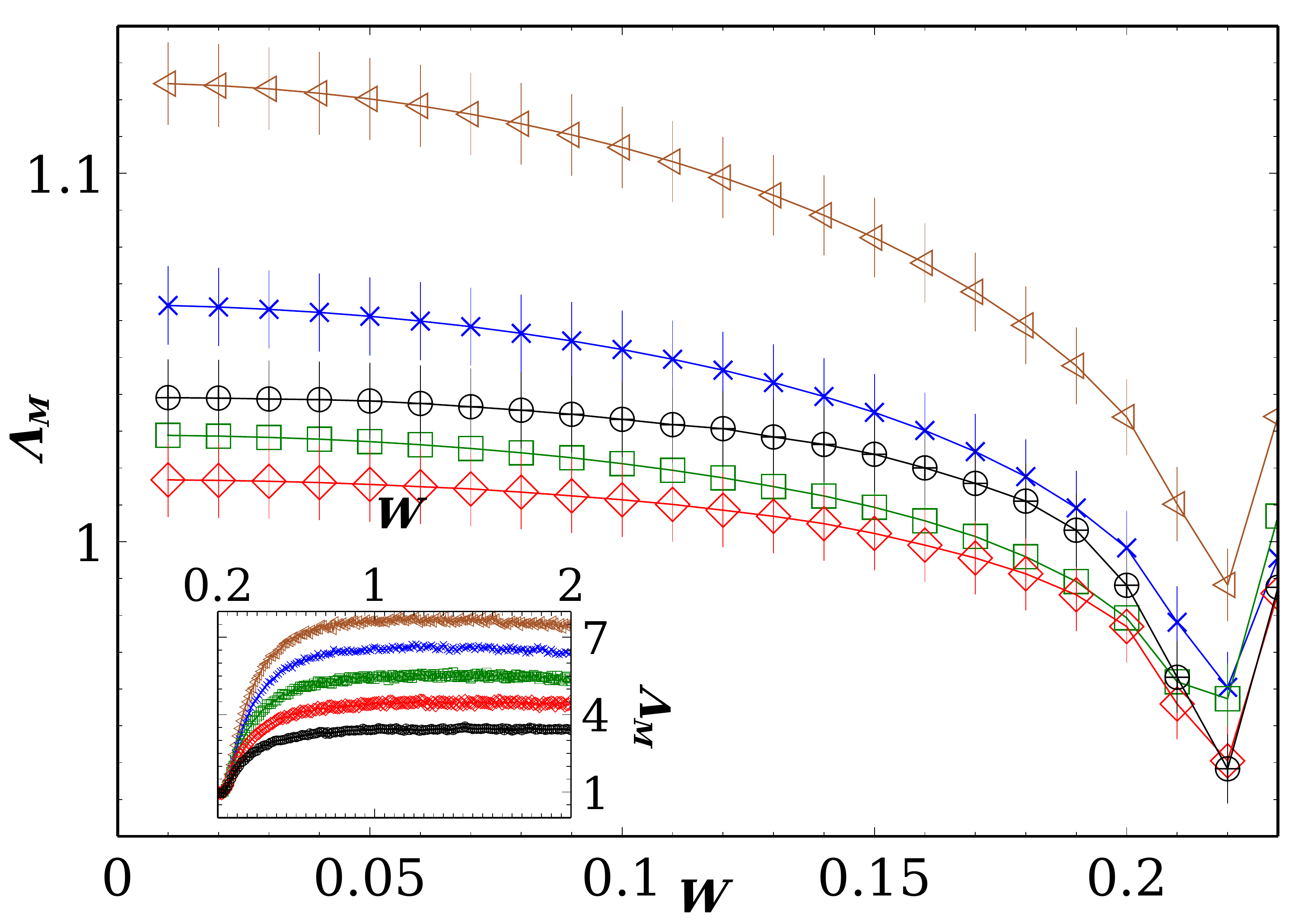}\\
    (c)\includegraphics[width=0.45\columnwidth]{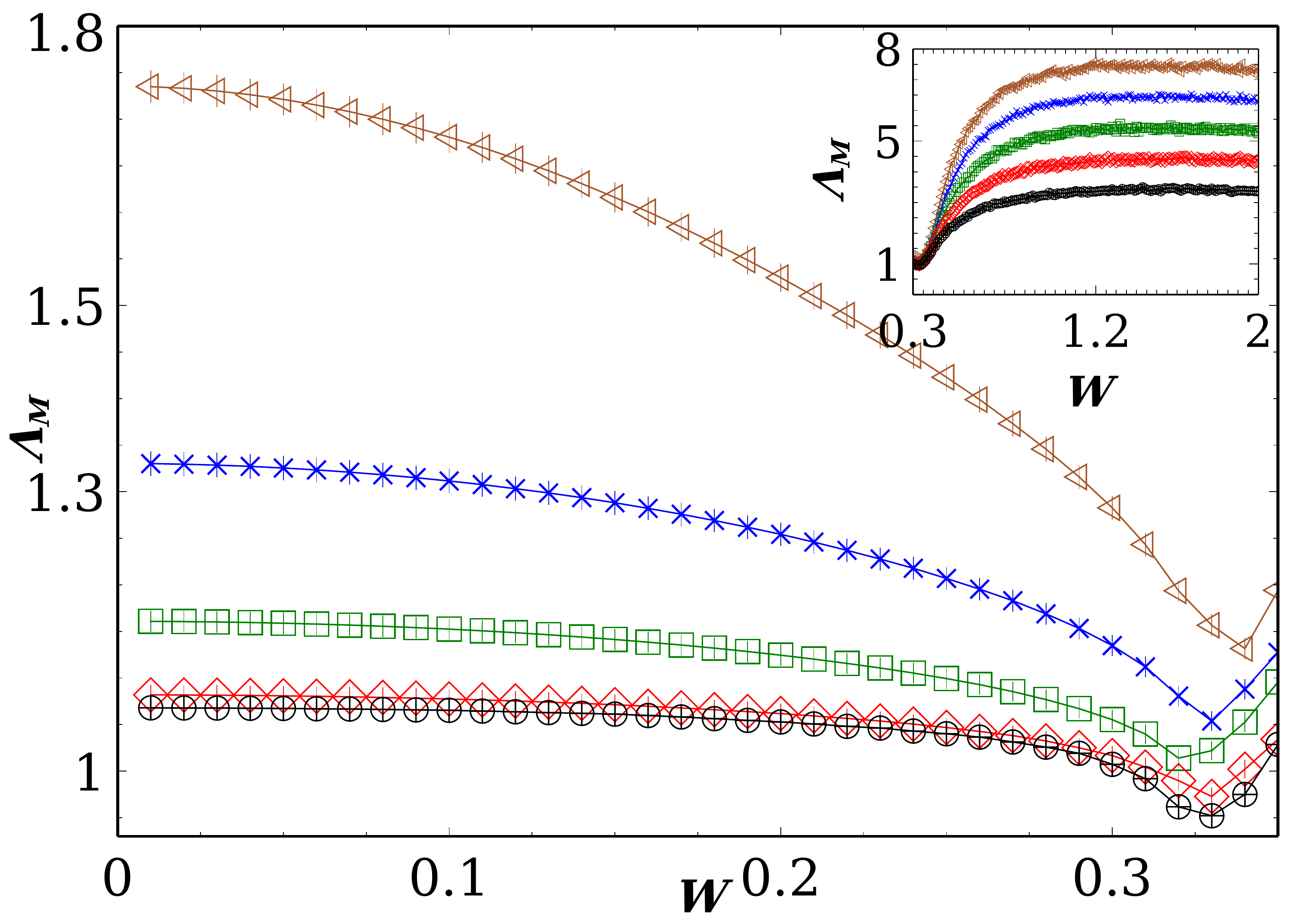}
    (d)\includegraphics[width=0.45\columnwidth]{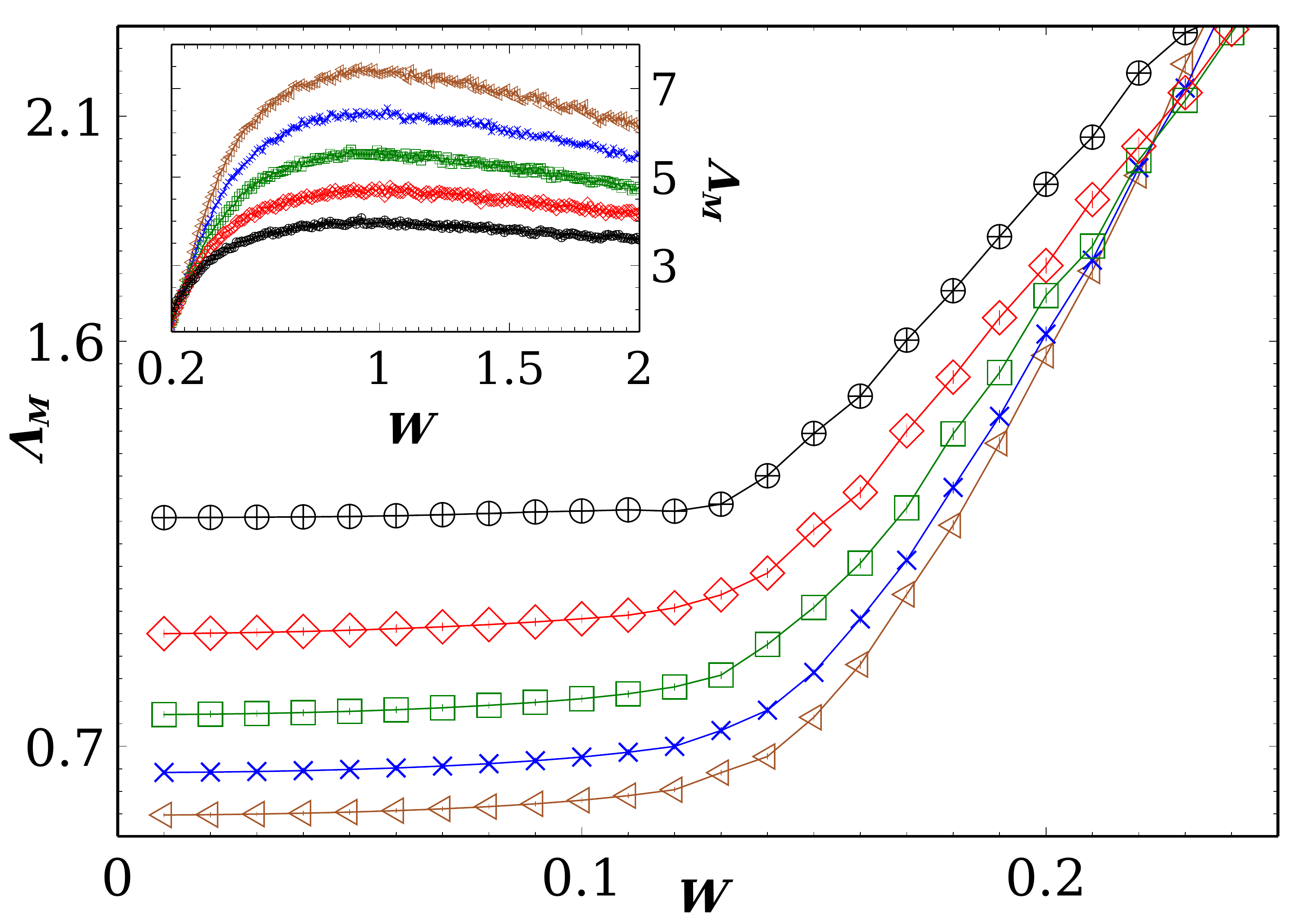}
    \caption{(a) Small $W$ behaviour of $\Lambda_M$ for $\mathcal{L}_3(1)$ with (a) energy $E=0.05$, (b) $E=0.1$, (c) $E=0.15$, and (d) for $\mathcal{L}_3(2)$ at $E=1.05$  with disorder down to $0.01$ in steps of $0.01$ and with error less than $1.0\%$. System sizes $M$ are $4$ (black $\oplus$), $6$ (red $\diamond$), $8$ (green $\square$), $10$ (dark-blue $\times$), $12$ (brown $\triangleleft$), $14$ (purple $\triangleright$). Error bars are denoted with a solid line. Insets: increased disorder range up to $W=2$ for the corresponding cases in the $4$ main panels.}
    \label{fig:Localization_Extended_transition}
\end{figure}

For $W<1$, it is well known that the convergence of the TMM is very slow. Hence results for appropriately small errors are hard to compute. Usually, this is not a problem since, e.g., in the 3D Anderson model, the limit as $W\rightarrow 0$ belongs trivially to the extended phase. However, for the $\mathcal{L}_3(n)$ lattice, we know that at the flat band energies even at $W=0$, we except compactly localized states \cite{Aoki1996HofstadterBands,Maimaiti2017CompactDimension}. Hence it is interesting to see if the localization properties at $W<1$ for flat band energies indicate any possible ``inverse'' Anderson transition from extended states at $W\sim 1$ to localized states at small, but finite $W>0$. 

In Ref.\ \cite{Liu2020LocalizationLattices}, we had shown that at the flat band energy $E=0$ for $\mathcal{L}_3(1)$ and at $E=1$ for $\mathcal{L}_3(2)$, the $\Lambda_M$ increases with increasing $M$, indicating extended behaviour, down to disorders as small as $W=0.01$. 
In Fig.\ \ref{fig:Localization_Extended_transition}(a-c), we now augment that result by studying energies close by. For $\mathcal{L}_3(1)$ and $E=0.05$ we initially find localized behaviour, e.g., $\Lambda_M$ decreasing with increasing $M$ up to $M=10$, but then reversing to extended behaviour for larger $M$. For $E=0.1$ the reversal to the extended behaviour already starts at $M=8$ while for $E=0.15$, only the extended behaviour remains. 
In Fig.\ \ref{fig:Localization_Extended_transition}(d), we see that for $\mathcal{L}_3(1)$ at $E=1.05$, the $\Lambda_M$ decreases with increasing $M$. This localized behaviour should vanish for larger $M$ values, but it is at present beyond our computational capabilities.

We conclude that the presence of the compactly localized states at the flat band energies certainly has an effect at small disorder, but for larger disorder values, the broadening of the flat bands and the mixing with the dispersive bands becomes dominant such that the character if the states is extended down to disorders $W=0.01$. This is usually already true at $W\sim 0.1$.

\subsection{Divergence of the scaling parameter $\xi(W)$}
The behaviour of $\xi(W)$ for $\mathcal{L}_3(1)$, $\mathcal{L}_3(2)$ and $\mathcal{L}_3(3)$ is given in Fig.\ \ref{fig:Scaling_Parameter_L3x}. We can clearly see how the critical disorder $W_c$ decreases from $8.59$ for $\mathcal{L}_3(1)$ to $5.96$ for $\mathcal{L}_3(2)$ and finally to $4.79$ for $\mathcal{L}_3(3)$. 
This suggests that a larger $n$ in $\mathcal{L}_3(n)$, i.e.\ a larger number of additional (red, cp.\ Fig.\ \ref{fig:Lieb_schematic}) atoms, leads to stronger localization and hence an MIT already for smaller values of $W_c$. It could be an interesting study to estimate $W_c(n)$, particularly the limiting behaviour when $n \rightarrow \infty$.


\subsection{Scaling with irrelevant variables $n_i$, $m_i$}

For high-precision estimates of critical properties, including $\nu$, it is by now state of the art to include irrelevant scaling contributions, i.e.\ scaling as $M^{-y}$ with $y>0$, in the FSS analysis. However, such FSS methods also require large $M$ values to reliably model the irrelevant scaling. Due to the complexity of the $\mathcal{L}_3(n)$ systems, only values of $M\leq 20$ have been computed in Ref.\ \cite{Liu2020LocalizationLattices}. For such sizes, adding irrelevant scaling variables is usually not a net benefit.
In Table \ref{table:critical parameters for L3x}, we show the results for FSS with and without scaling. We note that although acceptable $p$ values can be obtained for the fits with irrelevant scaling exponent $y$ included, in nearly all cases, this results either in increased error estimates for the relevant exponent $\nu$. Alternatively, one finds estimates for $y$ with very large errors or very large values for $y$. Except for one case, the final estimate for the physical quantity $\nu$ has hardly changed. Hence we conclude that for the available $\Lambda_M$ data, the inclusion of irrelevant scaling parameter $y$ does not necessarily add towards the accuracy of the estimates for $\nu$. This confirms having made this choice in Ref.\ \cite{Liu2020LocalizationLattices}.


\begin{table*}[tbh]
\centering
    \setlength{\tabcolsep}{0.0mm}{ 
    \begin{tabular}{cccccccccccccc}
    \hline \hline \noalign{\smallskip}
    \multicolumn{14}{c}{$\mathcal{L}_3(1)$}\\[0.7ex]
    $\Delta M$&   $E$&  $\delta W$&   $n_{r}$& $n_{i}$& $m_{r}$& $m_{i}$&   $W_{c}$&  CI($W_c$)& $\nu$&  CI($\nu$)&  $y$&  CI($y$)&  $p$    \\ 
    16-20&          0&    8.25-8.9&         3&       0&       1&       0&   $\textbf{\emph{8.59}}$&  $(58,61)$& $\textbf{\emph{1.6}}$& $(4,7)$&  0& 0& $0.15$     \\ 
    16-20&          0&    8.25-8.9&       2&     1&           1&      1&   ${{8.71}}$&  $(57,84)$&  ${{1.3}}$& $(0.8,1.8)$& 4& $(-2,10)$& $0.86$     \\[-2ex] 
\\
    14-20&          1&     8.0-8.8&        3&  0&  1&   0&         $\textbf{\emph{8.44}}$& $(42,45)$& $\textbf{\emph{1.6}}$& $(5,7)$& 0&  0&  $0.18$     \\ 
    14-20&          1&     8.0-8.8&        3&  2&  1&   1&           $8.48$&   $(45,50)$&    $1.8$&  $(6,9)$&  $6.9$& $(6.6,7.1)$&   $0.77$     \\ 

\\
    $\Delta M$&  $W$&  $\delta E$&    $n_{r}$& $n_{i}$& $m_{r}$& $m_{i}$&  $E_{c}$&  CI($E_c$)& $\nu$&  CI($\nu$)&  $y$&  CI($y$)& $p$    \\ 
    16-20&      3&    3.725-3.785&          2&  0&  1&  0&        $\textbf{\emph{3.75}}$& $(74,75)$& $\textbf{\emph{1.7}}$& $(6,9)$& 0&  0& $0.88$    \\ 
    16-20&      3&    3.725-3.785&          3&  2&  1&  2&        $3.75$&  $(74,75)$&  $1.5$&  $(0.6,2.5)$&   $2$& $(-3,8)$& $0.7$    \\[-2ex] 
    
\\
    16-20&           6&    3.04-3.11&       1& 0& 1& 0&          $\textbf{\emph{3.08}}$& $(07,09)$& $\textbf{\emph{1.5}}$& $(1.0,2.1)$& 0& 0& $0.14$  \\
    16-20&           6&    3.04-3.11&       1& 1& 2& 1&           $3.08$&   $(06,09)$&       $1.5$&  $(0.7,2.4)$&  $47$& $(44,50)$& $0.13$  \\
    
\\[0.5ex]\hline\noalign{\smallskip}
\multicolumn{14}{c}{$\mathcal{L}_3(2)$}\\[0.7ex]
    $\Delta M$ &    $E$&     $\delta W$&     $n_{r}$& $n_{i}$& $m_{r}$& $m_{i}$&    $W_{c}$&   CI($W_c$)& $\nu$&   CI($\nu$)&  $y$&  CI($y$)& $p$   \\ 
    12,14,18  &     0  &     5.85-6.05 &     2& 0&  2& 0& $\textbf{\emph{5.96}}$& $(95,97)$& $\textbf{\emph{1.8}}$& $(1.5,2.0)$&  0& 0&  $0.08$ \\
    12,14,18  &     0  &     5.85-6.05 &     2& 1&  1& 4& $5.97$& $(96,98)$& $1.7$& $(1.3,2.1)$&  $9$& $(2,16)$&  $0.89$ \\[-2ex]   
\\
    $\Delta M$&     $W$&    $\delta E$&   $n_{r}$& $n_{i}$& $m_{r}$& $m_{i}$&   $E_{c}$&  CI($E_c$)& $\nu$&  CI($\nu$)&   $y$&  CI($y$)&  $p$         \\ 
    10-14  &     4&      1.6-1.8 &            2& 0& 1& 0&          $\textbf{\emph{1.70}}$& $(70,71)$& $\textbf{\emph{1.6}}$& $(4,7)$& 0& 0& $0.18$ \\
    10-14  &     4&      1.6-1.8 &            1& 1& 2& 1&           $1.72$&   $(67,78)$&   $1.6$&  $(1.1,2.1)$&   $6$& $(-18,31)$&   $0.38$       \\

\\[0.5ex]\hline\noalign{\smallskip}
\multicolumn{14}{c}{$\mathcal{L}_3(3)$}\\[0.7ex]
    $\Delta M$&    $E$&      $\delta W$&      $n_{r}$& $n_{i}$& $m_{r}$& $m_{i}$&  $W_{c}$& CI($W_c$)& $\nu$& CI($\nu$)&   $y$&  CI($y$)& $p$     \\ 
    12-18     &     0 &     4.7--4.875 &        2& 0& 1& 0&      $\textbf{\emph{4.79}}$& $(78,80)$&  $\textbf{\emph{1.6}}$& $(4,8)$& 0& 0&  $0.43$  \\
    12-18     &     0 &     4.7--4.875 &        2& 1& 1& 2&      $4.79$& $(78,80)$&     $1.6$&  $(4,8)$&  $8284$& $(0,1)$&   $0.11$   \\
                                     \\
 \hline\hline

\end{tabular}
}
\caption{Critical parameters at the MIT for $\mathcal{L}_3(n)$, $n=1, 2$ and $3$. The columns are denoting the system width $M$, fixed $E$ (or $W$), the range of $W$ (or $E$). The expansion orders $n_{r}$, $n_{i}$, $m_{r}$, $m_{i}$ are listed as well as the obtained critical disorders $W_c$ (or energies $E_c$), their 95$\%$ confidence intervals (CI), the critical exponent $\nu$, its CI, the irrelevant parameter $y$, its CI, and the goodness of fit probability $p$. The confidence interval are given with one significant decimal. For instance, $1.6(4, 8)$ marks that the CI is $(1.4, 1.8)$
}
\label{table:critical parameters for L3x}
\label{tab:L32-3_E0000_FSS}
\end{table*}

\subsection{Density of states without Gaussian broadening for $\mathcal{L}_3(n)$}

The results of for the DOS, calculated with exact diagonalization and without applied Gaussian smoothing, are in Fig.\ \ref{fig:DOS without Gaussian smoothing}. The system sizes are $M^3= 5^3$, $5^3$, $4^3$, $4^3$ for $\mathcal{L}_3(n)$, $n=1, 2, 3$ and $4$, respectively. The disorder ranges are all from $W=0$ to $W=5.2$ in step of $0.05$ and with $300$ samples for $n=1, 2, 3$ but only $100$ samples for $\mathcal{L}_3(4)$ because of computing time limits. Again, the results are very similar to the Gaussian-broadened DOS shown in Ref.\ \cite{Liu2020LocalizationLattices}. 


\section{Conclusions}
\label{sec-conclusion}
We have studied the localization properties of the 2D and 3D extended Lieb lattices. Clearly, the Lieb lattices exhibit stronger localization than their more standard square and cubic Anderson lattices. This can be understood by noting that the transport along the "extra" sites as shown (by red spheres) in Fig.\ \ref{fig:Lieb_schematic} is essentially one-dimensional and hence subject to stronger localization. Consequently, in 2D rather small $W$ values can still be studied (most earlier TMM studies for the 2D Anderson model stop already around $W\sim 2$, cp.\ Fig.\ (3) of Ref.\ \cite{Leadbeater1999}). Details of the resulting FSS curves for small $W$ are given in Fig.\ \ref{fig:Scaling_Func_L2x}.
In 3D, we similarly see that $W_c(n)$ decreases as a function of $n$. Results for particularly small $W$ are shown in Fig.\ \ref{fig:Localization_Extended_transition}. Due to the numerical complexity of the $\mathcal{L}_d(n)$ systems, scaling is more challenging than in the Anderson models and only relatively small $M$ values can be reached. Table \ref{table:critical parameters for L3x} shows that for the available data, there is no need to include irrelevant scaling variables --- within the accuracy of the calculation, all estimates of the critical exponent agree with the value found for the Anderson universality class \cite{Slevin1999b,Rodriguez2011MultifractalTransition}.



\end{document}